\documentclass[aps,prb,twocolumn,superscriptaddress]{revtex4}

\usepackage{amsmath}
\usepackage{amsfonts}
\usepackage{amssymb}
\usepackage{graphicx}
\begin{document}


\title{The Kondo Effect in Ferromagnetic Atomic Contacts}


\author{M. Reyes Calvo}
\affiliation{Departamento de Fisica Aplicada, Facultad de Ciencias, Universidad de Alicante, San Vicente del Raspeig, E-03790 Alicante, Spain}

\author{Joaqu\'in Fern\'andez-Rossier}
\affiliation{Departamento de Fisica Aplicada, Facultad de Ciencias, Universidad de Alicante, San Vicente del Raspeig, E-03790 Alicante, Spain}

\author{Juan Jos\'e Palacios}
\affiliation{Departamento de Fisica Aplicada, Facultad de Ciencias, Universidad de Alicante, San Vicente del Raspeig, E-03790 Alicante, Spain}

\author{David Jacob}
\affiliation{Department of Physics and Astronomy, Rutgers University, Piscataway, New Jersey 08854, USA.}

\author{Douglas Natelson}
\affiliation{Department of Physics and Astronomy, Rice University, Houston, Texas 77005, USA.}

\author{Carlos Untiedt}
\thanks{Corresponding author: untiedt@ua.es}
\affiliation{Departamento de Fisica Aplicada, Facultad de Ciencias, Universidad de Alicante, San Vicente del Raspeig, E-03790 Alicante, Spain}



\begin{abstract}
Iron, cobalt and nickel are archetypal ferromagnetic metals. In bulk, electronic conduction in these materials takes place mainly
through the $s$ and $p$ electrons, whereas the magnetic moments are mostly in the narrow $d$-electron bands, where they tend to align.
This general picture may change at the nanoscale because electrons at the surfaces of materials experience interactions that differ from
those in the bulk. Here we show direct evidence for such changes: electronic transport in atomic-scale contacts of pure ferromagnets
(iron, cobalt and nickel), despite their strong bulk ferromagnetism, unexpectedly reveal Kondo physics, that is, the screening of local
magnetic moments by the conduction electrons below a characteristic temperature \cite{uno}. The Kondo effect creates a sharp resonance at
the Fermi energy, affecting the electrical properties of the system;this appears as a Fano-Kondo resonance \cite{dos} in the conductance
characteristics as observed in other artificial nanostructures\cite{tres,cuatro,cinco,seis,siete,ocho,nueve,diez,once}.
The study of hundreds of contacts shows material-dependent lognormal distributions of the resonance width that arise naturally
from Kondo theory \cite{doce}. These resonances broaden and disappear with increasing temperature, also as in standard Kondo systems
\cite{cuatro,cinco,seis,siete}. Our observations, supported by calculations, imply that coordination
changes can significantly modify magnetism at the nanoscale. Therefore, in addition to standard micromagnetic physics, strong
electronic correlations along with atomic-scale geometry need to be considered when investigating the magnetic properties of magnetic
nanostructures.
\end{abstract}


\maketitle

Atomic-scale contacts can be fabricated by techniques such as scanning tunnelling microscopy \cite{trece}
or the use of electromigrated break junctions (EBJs)\cite{catorce}, where the size of a macroscopic contact between
two leads is reduced until they are in contact through only a few atoms and, eventually, through only one. 
The conductance of metallic monatomic contacts is known to be around $2G_{0}$, where $G_{0}=e^2/h$ is the spin-resolved 
quantum of conductance\cite{trece} ($e$ being the elementary charge and $h$ Planck$'$s constant). To identify the atomic contacts,
histograms are constructed from the evolution of the conductance recorded during the breaking of different contacts (Fig. 1a, b). The
position of the first peak of these histograms is identified as the conductance of the monatomic contact. For iron, cobalt and nickel, 
the conductance is larger than $2G_{0}$ owing to the contribution of the $sp$ and $d$ orbitals to the transmission \cite{quince,dieciseis,
diecisiete}. 

\begin{figure*}[ht]
\includegraphics[width=0.9\linewidth]{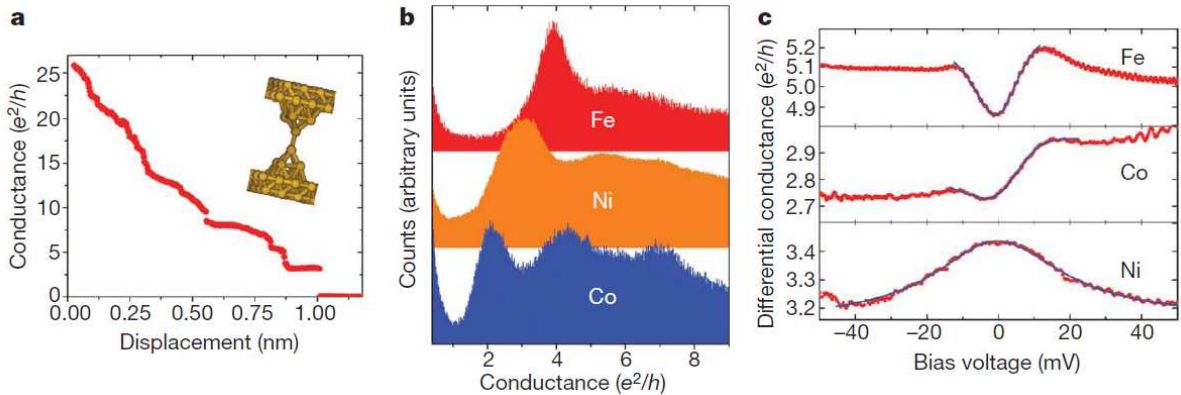}
\caption{Conductance of a monatomic contact. \textbf{a}, Example of a trace where we record the conductance while stretching a 
nickel wire using a scanning tunnelling microscope (STM) at 4.2 K. Inset, model of a monatomic contact.
\textbf{b}, Conductance histograms constructed for iron, cobalt and nickel from thousands of such traces. 
The position of the first peak of in each histogram corresponds to the conductance of the monatomic contact. \textbf{c}, Differential
conductance curves recorded at the monatomic contact as a function of the applied voltage. A characteristic resonance appears at 
small bias that fits the Fano line shape. All possible symmetries are found in the spectroscopy of iron, cobalt and nickel contacts, 
and the width of the resonance is the main difference between the spectra of the three materials. This width is proportional to the
Kondo temperature.}
\end{figure*}

We have studied the low-temperature conductance characteristics of hundreds of atomic-scale contacts of the three transition-metal
ferromagnets iron, cobalt and nickel using a home-built STM. More than the 80\% of the differential conductance (d$I$/d$V$) curves at the
monatomic contact show peaks or dips around zero bias such as those shown in Fig. 1c, which are very similar to those observed in
STM spectroscopy of single magnetic adatoms on non-magnetic surfaces\cite{nueve,diez,once}. Thus, as in the case of these Kondo systems, we can also
fit our d$I$/d$V$ curves to the sum of a flat component, $g_{0}$, and a Fano-like resonance that typically amounts for 10\% of the signal:

\begin{equation}
\frac{dI}{dV}=g_{0}+\frac{A}{1+q^2}\frac{(q+\epsilon)^2}{1+\epsilon^2} 
\end{equation}

Here $\epsilon=(eV-\epsilon_{s})/k_{B}T_{K}$ is the bias shifted with respect to the centre of the resonance, $\epsilon_{s}$, and
normalized by the natural width of the resonance, $k_{B}T_{K}$; $T_{K}$ is the Kondo temperature; $q$ is the dimensionless Fano parameter
that determines de simmetry of the curve; $A$ is the amplitude of the feature; and $k_{B}$ is Boltzmann's constant.


It is clear from the d$I$/d$V$ characteristics in Fig. 1c that the width of the Fano feature is different for each material. 
This is confirmed by statistical analysis of the data. Figure 2 shows histograms of $T_{K}$ obtained from fitting our conductance 
curves for hundreds of different iron, cobalt and nickel monatomic contacts. Notably, these histograms follow a log-normal distribution;
that is, the natural logarithm of $T_{K}$ is normally distributed. Because many different atomic configurations
result in monatomic contacts, their electronic properties, such as conductance (Fig. 1b), density of states and the associated 
energy scales, are expected to be normally distributed. Instead, a normal distribution of
$log(T_{K})$ can only be understood if $T_{K}$ can be expressed as the exponential
of normally distributed quantities. The Kondo model\cite{doce} naturally relates $T_{K}$ to the exponential of the typical
energy scales in the problem.

Fitting the histograms to a log-normal distribution yields most frequent values for the resonance widths in the different materials
corresponding to $T_{K}$=90 K, 120 K and 280 K for iron, cobalt and nickel, respectively, following the same trend ($T^{Fe}_{K} <
T^{Co}_{K}<T^{Ni}_{K}$) for these chemical species when deposited as adatoms on non-magnetic
surfaces\cite{dieciocho}. In simple terms, the Kondo temperature decreases as the size of the screened magnetic moment 
increases, as we change from nickel to cobalt to iron. The Kondo temperature of cobalt nanocontacts
is very similar to the one reported in ref. \cite{once} for cobalt in Cu(100) probed using a STM in the contact regime.

\begin{figure}
\includegraphics[width=0.9\linewidth]{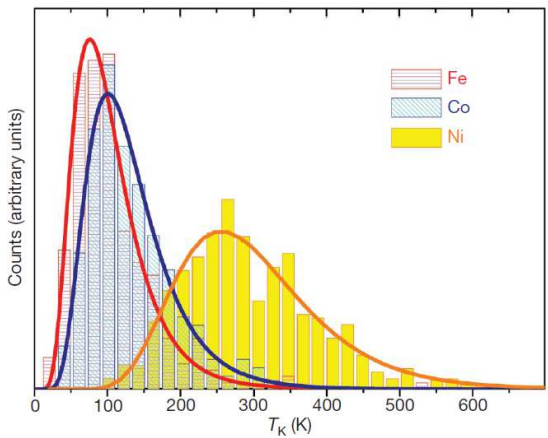}
\caption{Histograms of inferred Kondo temperatures for iron, cobalt and
nickel. The histograms are constructed from more than 200 fittings and
normalized to the total number of curves fitted. The continuous lines show
the fits of the data to log-normal distributions of $T_{K}$ with a different most
probable value for each material.}
\end{figure}

In addition to the statistical analysis described above, we measured the temperature evolution of the Kondo features for a 
single contact. To preserve the atomic stability of the junction while changing the temperature \cite{diecinueve}, 
we used an EBJ such as the one shown in the inset of Fig. 3b. It is important to note that in this alternative implementation
we observe exactly the same Fano resonances, with the same distribution of Kondo temperatures, as we obtained in the STM experiments
using nickel and cobalt samples. In all the cases, we see a reduction of the amplitude of the Fano features as the temperature
increases, as shown in Fig. 3a for the case of cobalt. In Fig. 3b we plot $A(T)$, as defined in equation (1), as a function of 
$T$ on a logarithmic scale. The curve so obtained has a low-temperature plateau followed
by a linear decay, very similar to that of quantum dots and molecules in the Kondo regime \cite{cuatro,cinco,seis,veinte}.

In summary, our atomic contacts fabricated with two different methods show the same d$I$/d$V$ curves (Fig. 1c), the same chemical
trends (Fig. 2) and the same temperature evolution (Fig. 3a, b) as other standard, chemically inhomogeneous Kondo systems. 
This indicates that the contact atom(s) in nanocontacts of iron, cobalt and nickel are in the Kondo regime. 
This is unexpected for two reasons. First, the Kondo effect has always been associated with chemically inhomogeneous
systems containing at least two kinds of atom: those where the localized level resides and those providing the itinerant electrons.
Here the same chemical species hosts both the itinerant states and the local magnetic moments. Second, in most reports on the Kondo
effect, the itinerant electrons are not spin-polarized, as a large spin polarization is expected to destroy the effect. 
or a carbon nanotube \cite{vintidos} contacted with ferromagnetic leads has been justified by the oppositely directed spin polarizations in the electrodes,
in agreement with refs. \cite{vintitres,vinticuatro,vinticinco}. Although the situation regarding the
magnetization of the leads might be similar in our system, we argue that the Kondo effect is still possible even if there is 
no domain wall pinned in the contact.

\begin{figure}
 \includegraphics[width=0.9\linewidth]{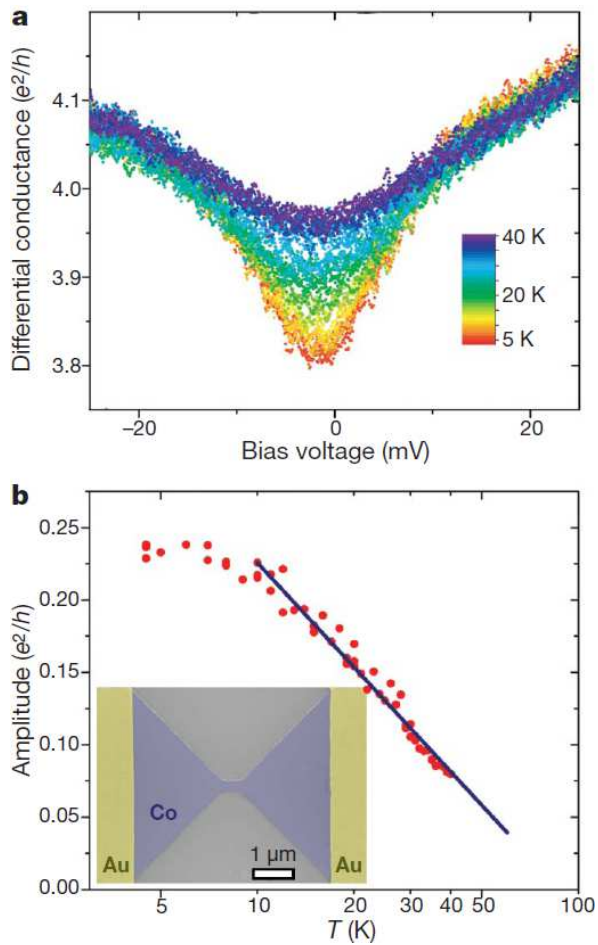}
\caption{Evolution of the Fano resonances with increasing temperature.
\textbf{a}, Characteristics of differential conductance versus bias voltage for a cobalt
atomic contact, showing how the Fano resonance disappears as the
temperature is increased. \textbf{b}, The amplitude of the Kondo resonance from
a decreases logarithmically with the temperature. Inset, an artificially
coloured example of a lithographic device for the EBJ experiments (before
electromigration); the cobalt junction is blue. The atomic contacts created
by this method are suitable for studying the temperature dependence of the
Fano resonance.}
\end{figure}

Being in the Kondo regime implies that the atoms at the contact must host, at least, a localized $d$-electron level whose magnetic
moment is screened as a result of antiferromagnetic coupling to the $sp$ conduction electrons. We can show how local moment formation
and antiferromagnetic coupling occur in nickel nanocontacts in the following way (the cases of iron and cobalt can be understood on
similar grounds). The mean-field solution of the Anderson model \cite{vintiseis}
,which describes a localized $d$ level, with energy $\epsilon_{d}$, on-site repulsion $U$ and hybridization with the itinerant
$sp$ electrons, $V_{sp-d}$, defines the conditions for the formation of a local moment in such a $d$ level. The
model can also be used to derive the antiferromagnetic $sd$ exchange coupling, $J^{AF}_{sd}$ (ref. \cite{vintisiete}
), which arises from the $sp-d$ hybridization term. Here we used the local spin-density approximation (LSDA) to density
functional theory and its generalization LSDA+U as the mean field to determine the formation of a local moment and its 
antiferromagnetic coupling to the $sp$ carriers. We consider both monostrand chains and nanocontacts. 
Their smaller atomic coordination, compared with that of the bulk, results in a stronger electronic localization and a larger
magnetic moment per atom (values of 1.17 Bohr magnetons, compared with 0.6 Bohr magnetons in the bulk, were obtained using
LSDA). This favours the appearance of the Kondo effect. In Fig. 4a, b, we show the LSDA bands obtained for the nickel chain. 
Out of the six minority-spin bands crossing the Fermi level, the two degenerate E2 bands are the narrowest, 
hosting highly localized electrons. These are the bands that are less well described by the LSDA because of the
inherent self-interaction problem of this approximation. 

\begin{figure}
\includegraphics[width=0.9\linewidth]{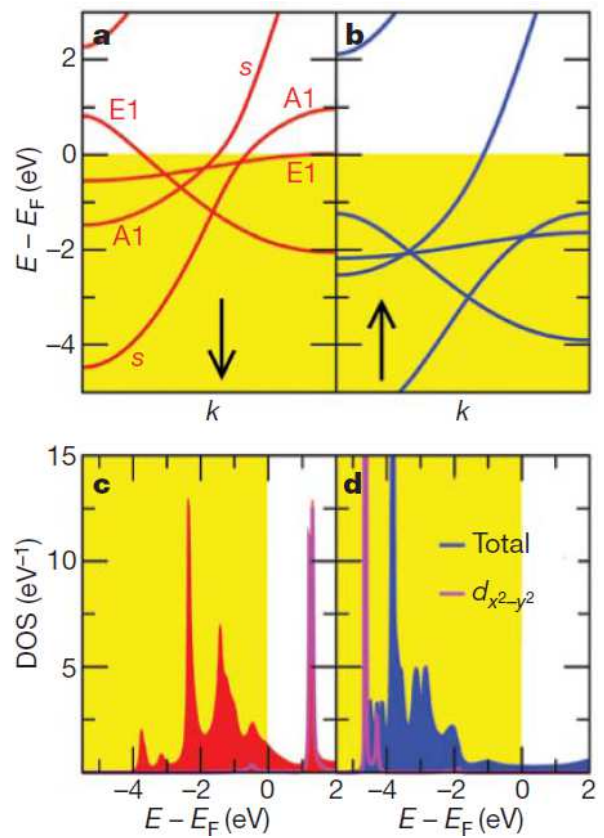}
\caption{Electronic structure for a nickel chain and a nickel nanocontact.
\textbf{a}, LSDA minority-spin energy bands for a monostrand nickel chain, where
the wavevector $k$ runs over the first Brillouin zone. The bands are labelled
according to their symmetry group: the E1 and E2 bands ($d$-band like,
doubly degenerate and decoupled from the $s$ band) and the A1 or $d_{3z^2-r^2}$
band (hybridized with the $s$ band). The lattice constant is 2.09 \AA\ . \textbf{b}, As in
\textbf{a}, but for majority-spin electrons. The yellow background indicates
occupied states. \textbf{c}, LSDA+U minority-spin results for the total DOS and the
DOS projected on the $d_{x^2-y^2}$ orbital for a tip atom of a nickel nanocontact,
with $U$= 3 eV. \textbf{d}, As in \textbf{c}, but for majority-spin electrons.}
\end{figure}

We modelled actual nanocontacts with two identical pyramids
facing each other in the $[001]$ direction. As previously done for chains\cite{vintiocho}
, and to avoid the self-interaction problem, we computed the electronic structure of the nanocontact using LSDA+U. In
Fig. 4c, d, we show the total density of states (DOS) projected on a tip atom and the DOS projected on the corresponding
$d_{x^2-y^2}$ orbitals for both the minority (Fig. 4c) and the majority (Fig. 4d) spins (here
we have set $U$=3 eV). In general, solutions obtained for values of $U$ ranging from 3 to 5 eV also show that 
the $d_{x^2-y^2}$ orbital, forming one
of the E2 bands in chains, hosts an integer local magnetic moment.
Different geometries may favour the formation of the local moment
in other strongly localized orbitals. The hybridization of the $d_{x^2-y^2}$
orbital with the surrounding $sp$ orbitals results in antiferromagnetic kinetic exchange \cite{vintisiete}
. We estimate that $J^{AF}_{sp-d} \simeq$ 1 eV, taking $\epsilon_{d}\simeq$ 5 eV,
$|V_{sp-d}|^{2}\simeq$ 2 eV and $U_{eff}\simeq$ 8 eV from our LSDA+U calculations,
where $U_{eff}$ is the spin splitting of the $d_{x^2-y^2}$ orbital (Supplementary
Information). We note that, in contrast to the contact atom(s), the intra-atomic
$sp-d$ hybridization vanishes for bulk atoms owing to the very small
anisotropy of the crystal environment. This makes the antiferromagnetic
coupling $J^{AF}_{sp-d}$ larger in the contact than in the bulk. This coupling
competes with the ferromagnetic coupling $J^{FM}_{sp-d}$, which we
estimate from the splitting of the $sp$ band at the boundary of the
Brillouin zone for chains to be $\simeq$0.2 eV (Fig. 4a, b). Thus, an overall
antiferromagnetic coupling between $d$ electrons and $sp$ conduction
electrons is possible in nickel nanocontacts. Additionally, the local
moment, \textbf{m}, responsible for the Kondo effect is subject to the ferromagnetic
coupling $J_{dd}$ to the neighbouring atoms, and this interaction
also competes with the antiferromagnetic $sp-d$ coupling:

\begin{equation}
H_{exch} = \bf{m} \cdot \left[(J^{AF}_{sp-d}-J^{FM}_{sp-d})\bf{S_{s}}-J_{dd} \sum_{i} \bf{m}_{i}  \right]
\end{equation}

Here the \textbf{m}$_i$ are the local moments of the neighbouring atoms and $\bf{S_s}$
is the spin of the $s$ electrons. In ref. \cite{vintinueve}
, the coupling $J_{dd}$ was calculated for iron, cobalt and nickel by implementing the magnetic force
theorem with a LSDA ground state. This method yields values for the spin-wave dispersion of the materials that compare well with
experiment \cite{treinta}
, and gives $J_{dd}$=19 meV, 15 meV and 2.7 meV for iron,
cobalt and nickel, respectively. Thus, $J_{dd}$ is significantly smaller than
$J^{AF}_{sp-d}$. 

The Kondo effect in nanocontacts is favoured by three factors. First,
a local moment forms in the contact atoms because of their smaller
coordination. Second, the reduced symmetry of the contact,
compared with the bulk, enhances the intra-atomic contribution to
the $sp-d$ hybridization and, thus, the antiferromagnetic coupling
$J^{AF}_{sp-d}$ . Third, the smaller coordination also reduces the influence of
the direct ferromagnetic $dd$ coupling with neighbouring atoms. As a
result, the local moment formed in the contact is antiferromagnetically
coupled to the sp itinerant electrons and results in the Kondo
effect in this system, in contradiction to conventional wisdom.

\begin{acknowledgments}
We thank to E. Tosatti, R. Aguado and J. Ferrer for discussions, G. Scott and G. Saenz-Arce for experimental support and V. Esteve
for technical support. This work was partly supported by the European Union through MolSpinIQ and Spanish MEC (grants MAT2007-65487,
31099-E and CONSOLIDER CSD2007-0010). D.J. acknowledges funding by the US National Science Foundation (NSF) under grant DMR-0528969.
D.N. acknowledges the support of NSF grant DMR-0347253, the David and Lucille Packard Foundation and the W.M. Keck Program in Quantum
Materials.
\end{acknowledgments}

\appendix
\section{Methods}

For the statistical results on the Kondo parameters, we used a home-made STM
at 4.2K to fabricate the contacts by indentation between two pieces of metal \cite{trece}.
For the temperature-dependence measurements, the contacts were produced by
the controlled electromigration at 4.2K of 100-nm-wide junctions fabricated by
electron beam lithography \cite{diecinueve}. In both cases, the spectroscopic curves were
obtained by the addition of an a.c. voltage with a peak-power amplitude of
1mV and a frequency of 1 kHz to the d.c. bias voltage to allow the lock-in
detection of the differential conductance.

\paragraph{Fabrication of atomic-scale contacts by scanning tunnelling microscopy.}

Two pieces of metal wire of 0.1-mm diameter were cleaned and sonicated in acetone and isopropanol. 
These pieces were mounted in a home-built STM. The set-up was pumped down to high vacuum and immersed in 
a liquid helium bath until the temperature reached 4.2 K. The two pieces of wire were brought into contact 
and then pulled apart until the contact was of atomic dimension and, finally, until only one atom formed the contact\cite{trece}. 
Strong indentation before the formation of every contact ensured the cleanliness of the atomic contacts. 
The histograms obtained by this technique (Fig. 1a) show similar results to the ones obtained by 
the mechanically controlled break junction technique, where the surfaces brought into contact are created under cryogenic conditions.

\paragraph{Fabrication of atomic-scale contacts using EBJs.}

As a first step, a small junction about 100 nm wide was fabricated from cobalt using electron beam lithography and 
electron beam evaporation over a silicon dioxide substrate. To make the junction suitable for contact by macroscopic probes, 
two gold electrodes are deposited over the edges of the junction in a second lithography step, following the procedure described
in ref. \cite{diecinueve}. The samples are then placed in a probe station that was pumped down and immersed in a liquid helium
bath until the sample reached a temperature close to 4.2 K. Under these conditions, the controlled electromigration process 
\cite{catorce,diecinueve} was performed, decreasing the size of the junction to the atomic scale.

\section{Supplementary information on LSDA and LSDA+U results}

The electronic structure of a Ni nanocontact model as that shown in Fig. 5 has been calculated with density functional 
theory in the local spin-density approximation (LSDA)\cite{treintayuno}
using our established ab-initio transport methodology \cite{dieciseis}
. Fig. 5a shows the DOS and partial $d$ DOS for one of the tip atoms of the nanocontact. 
The majority-spin $d$-levels are completely filled while the minority-spin $d$-levels are partially occupied with an 
average occupation of 0.8. This results in a net magnetic moment of $\mu \simeq$1 for the tip atom similar to the magnetic moment
of the one-dimensional chain. 

It is, however, well known that LSDA suffers from the so-called self-interaction problem which blue-shifts occupied levels. 
Here the spurious self-interaction brings all the minority $d$-orbitals up to the Fermi level leading to a partial and almost
 equal occupation of 0.8 of all five $d$-orbitals. The self-interaction error of LSDA is corrected, e.g., in the LSDA+U method
\cite{treintaydos}
. In the LSDA+U method the effective LSDA Kohn-Sham potential of the strongly interacting $d$-electrons is
corrected by adding a Hartree-Fock term for an on-site Coulomb repulsion $U$ and exchange interaction $J$:

\begin{equation}
 \langle i \sigma | V_{LSDA+U} |i \sigma \rangle = U (N_{d}-n_{i}^{\sigma}) - J(N_{d}^{\sigma}-n_{i}^{\sigma}) -E_{dc}
\end{equation}

Here, $N_d$ is the total occupation of all the $d$-levels of an atom while $n_{i}^{\sigma}$ is the occupation of an individual
$d$-level per spin $\sigma$, and $N_{d}^{\sigma}$ is the total occupation of $d$-levels per spin $\sigma$. $E_{dc}$ accounts 
for the fact that the Coulomb repulsion and exchange interaction have already been taken into account in some way at the LSDA level,
and thus have to be subtracted from the effective LSDA Kohn-Sham potential in order to avoid double-counting \cite{treintaydos}
.

Due to the low coordination of the two tip atoms (compared to the rest of the nanocontact and to bulk atoms) Coulomb interaction 
effects are expected to be much stronger for the tip atoms than for the rest of the nanocontact. Hence, we only treat the two 
tip atoms of the nanocontact at the LSDA+U level while the rest of the nanocontact is treated at the LSDA level. 
For bulk Ni a widely accepted value is $U$=3 eV for the direct Coulomb repulsion and $J$=1eV for the exchange interaction 
\cite{treintaytres}
.
Due to the lower coordination of the tip atoms with respect to bulk atoms, the screening of the Coulomb repulsion should be 
lower than in bulk. Therefore the Coulomb repulsion is expected to be bigger than $U$=3 eV for the tip atoms. 

\begin{figure*}[htp]
\includegraphics[width=0.75\linewidth]{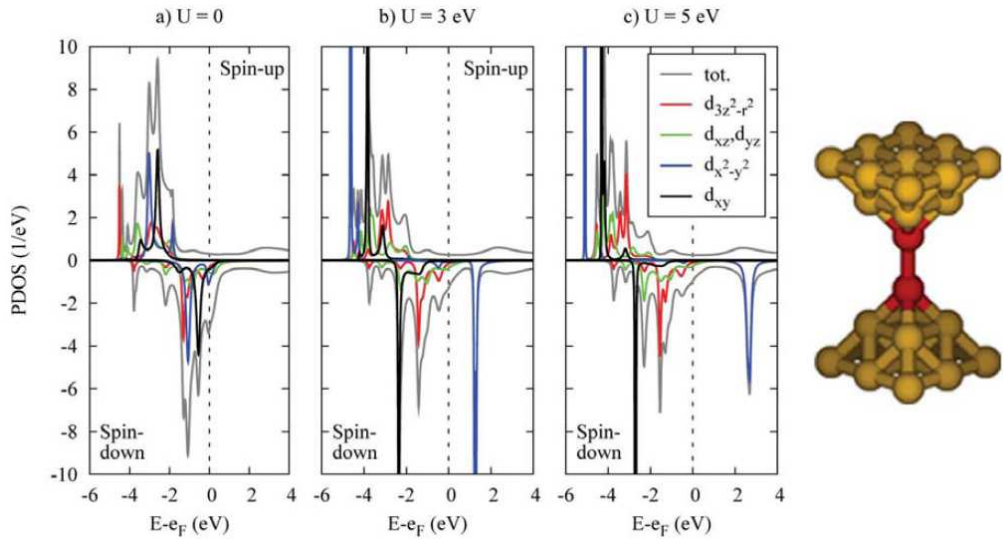} 
\caption{Total DOS (grey) and partial DOS (PDOS) for $d$-orbitals (red, green, blue, black) for the tip-atom of a Ni nanocontact 
calculated with the LSDA (a) and the LSDA+U method for different values of $U$ (b),(c). The Fermi level $E_{F}$ is set to zero. 
The $d_{xz}$ and $d_{yz}$-orbitals (green) are degenerate. For $U\geq$ 3 eV the $d_{x^2-y^2}$-orbital (blues) becomes completely 
spin-polarized while all other $d$-orbitals become doubly-occupied. (See text for further explanations). d)Model geometry of 
a Ni nanocontact. Two perfect pyramidal tips facing each other in the $[$001$]$ direction. Distance between the two tip atoms 
(in red) is 2.6 \AA\. Nearest neighbour distance between atoms in each of the pyramids is the same as in bulk Ni (2.49 \AA\ )}
\end{figure*}

Indeed, as can be seen from Figs. 5a,b, for $U \geq$3 eV four of the minority $d$-orbitals are now well below the Fermi energy 
and completely occupied, while the minority $d_{x^2-y^2}$−orbital is completely empty. Thus the magnetic moment of the tip atom 
is now entirely carried by the $d_{x^2-y^2}$-level. This shows that a local moment can form on the tip atom which is carried 
by a single $d$-orbital. The screening of the magnetic moment of this $d$-orbital by the $sp$-conduction electrons can thus give 
rise to a Kondo resonance at the Fermi level.

\begin{table}
 \begin{tabular}{c|c|c|c|c|c}
  U&	$|t_{sd}|$&	$|t_{pd}|$&	$\epsilon_{d}$& 	$\epsilon_{d}+\tilde{U}$&	$J_{sp-d}^{AF}$ \\ \hline
3 eV&	0.3 eV&		0.6 eV &	-4.5 eV&		1.5 eV &			1.6eV\\ \hline
5 eV&	0.3 eV&		0.6 eV &	-5 eV&			3 eV &				1.0eV\\ 
 \end{tabular}
\caption{Table of hoppings and excitation energies extracted from an LSDA+U calculation with $U$= 3 eV and $U$=5 eV, and $J$=1 eV
for all $d$-orbitals on the two tip atoms, as well as the resulting antiferromagnetic exchange coupling $J_{sp-d}^{AF}$.}
\end{table}

Now we estimate the exchange coupling $J^{AF}_{sp-d}$ of the $d_{x^2-y^2}$−orbital 
of the tip atoms from the hoppings and excitation energies of the effective Hamiltonian of the LSDA+U calculation. 
The $d_{x^2-y^2}$-orbital of the tip atom of one of the pyramids is coupled via hoppings $t_{sd}$ and $t_{pd}$ to the $s$ and 
$p_z$-orbitals of the four atoms next to the tip atom in the same pyramid as detailed in Tab. 1. The total antiferromagnetic 
exchange coupling is given by:

\begin{equation}
 J^{AF}_{sp-d}=|V_{sp-d}|^2 \left(\frac{1}{|\epsilon_{d}|}+\frac{1}{\epsilon_{d}+U_{eff}} \right)
\end{equation}

where $V_{sp-d}$ is the effective total hybridization with the $sp$-channels:

\begin{equation}
 |V_{sp-d}|^2=4|t_{sd}|^2+4|t_{pd}|^2
\end{equation}


\bibliography{nature_latex}

\end{document}